# Observation of bulk nodal lines in topological semimetal ZrSiS

B.-B. Fu,[1,2,#] C.-J. Yi,[1,2,#] T.-T. Zhang,[1,2,#] M. Caputo,[3] J.-Z. Ma,[1,2] X. Gao,[1,2] B. Q. Lv,[1,2] L.-Y. Kong,[1,2] Y.-B. Huang,[4] M. Shi,[3] V. N. Strocov,[3] C. Fang,[1] H.-M. Weng,[1,5,*] Y.-G. Shi,[1,*] T. Qian,[1,5,*] and H. Ding[1,2,5]

[1] *Beijing National Laboratory for Condensed Matter Physics and Institute of Physics, Chinese Academy of Sciences, Beijing 100190, China*

[2] *University of Chinese Academy of Sciences, Beijing 100049, China*

[3] *Paul Scherrer Institute, Swiss Light Source, CH-5232 Villigen PSI, Switzerland*

[4] *Shanghai Synchrotron Radiation Facility, Shanghai Institute of Applied Physics, Chinese Academy of Sciences, Shanghai 201204, China*

[5] *Collaborative Innovation Centre of Quantum Matter, Beijing, China*

[#] These authors contributed to this work equally.

[*] Corresponding authors: hmweng@iphy.ac.cn, ygshi@iphy.ac.cn, tqian@iphy.ac.cn

# Abstract

ZrSiS is the most intensively studied topological nodal-line semimetal candidate, which is proposed to host multiple nodal lines in its bulk electronic structure. However, previous angle-resolved photoemission spectroscopy (ARPES) experiments with vacuum ultraviolet lights mainly probed the surface states. Here using bulk-sensitive soft X-ray ARPES, we acquire the bulk electronic states of ZrSiS without any interference from surface states. Our results clearly show two groups of three-dimensional bulk nodal lines located on high-symmetry planes and along high-symmetry lines in the bulk Brillouin zone, respectively. The nodal lines on high-symmetry planes are enforced to pin at the Fermi level by carrier compensation and constitute the whole Fermi surfaces. This means that the carriers in ZrSiS are entirely contributed by nodal-line fermions, suggesting that ZrSiS is a remarkable platform for studying physical properties related to nodal lines.

Topological semimetals (TSMs) are defined as the systems whose valence and conduction bands cross each other. In Dirac and Weyl semimetals, the crossing of two doubly- or singly-degenerate bands forms 4- or 2-fold degenerate nodal points. Such Dirac and Weyl semimetals have been theoretically predicted [1-6] and experimentally confirmed [7-12]. Recently, band theory has predicted 3-, 6- and 8-fold degenerate nodal points [13-17], where the quasiparticles have no counterparts in high-energy physics, and the 3-fold degenerated points have been confirmed experimentally very recently [18,19]. By contrast, the band crossings in topological nodal-line semimetals (TNLSMs) extend along 1D line in the Brillouin zone (BZ). The TNLSM state is a versatile ground for other topological quantum states. Breaking certain symmetry can drive it to topological insulator, Weyl semimetal, or Dirac semimetal [20]. In contrast to the discrete nodal points, nodal lines have much richer topological configurations and they can constitute nodal chain, nodal link, or nodal knot [21-25].

So far, band structure calculations have predicted numerous TNLSM candidates, in which the nodal lines are protected by various symmetries [20,26-28]. Among them, PbTaSe$_2$, CaAgAs, and the *WHM* family, which represents a large number of materials with *W* = Zr, Hf, or La; *H* = Si, Ge, Sn, or Sb; *M* = O, S, Se, Te [29], have been studied by angle-resolved photoemission spectroscopy (ARPES) [30-39]. The measured samples of CaAgAs suffer from serious hole doping so that the proposed nodal line lies far above the chemical potential and thus cannot be observed by ARPES [37-39]. For PbTaSe$_2$, the bulk states are seriously interfered by surface states, causing difficulty to identify the bulk nodal lines [36]. The *WHM* materials, especially ZrSiS, have also been intensively studied by ARPES with vacuum ultraviolet (VUV) light. However, the VUV ARPES mainly probes the electronic states on the surface [30-35], which tremendously hinders exploring the intrinsic bulk electronic states.

In this work, we use bulk-sensitive soft X-ray ARPES to investigate the electronic structure of ZrSiS, the most intensively studied TNLSM candidate. The three-dimensional (3D) bulk nodal lines are clearly resolved without any interference from surface states. Furthermore, our results show that the whole Fermi surfaces are only composed of the nodal lines on high-symmetry planes, which provides an ideal playground for studying physical properties of nodal lines, especially in transport measurements.

High-quality single crystals used in the ARPES measurements were grown by chemical vapor transport method with iodine as agent. Soft X-ray ARPES measurements were performed at the Advanced Resonant Spectroscopies (ADRESS) beamline at the Swiss Light Source (SLS) [40] and the 'Dreamline' beamline of the Shanghai Synchrotron Radiation Facility (SSRF). Most of the soft X-ray ARPES data were taken at the ADRESS beamline with the overall energy resolution from 40 to 100 meV. VUV ARPES measurements were performed at the 'Dreamline' beamline and the 13U beamline of the National Synchrotron Radiation Laboratory (NSRL) at Hefei. All the samples were cleaved *in- situ* in vacuum condition better than $5 \times 10^{-11}$ Torr and measured at 20 K. First-principles calculations based on density functional theory (DFT) [41] were performed within the Perdew-Burke-Ernzerhof (PBE) exchange-correlation [42] implementing in the Vienna *ab initio* simulation package (VASP) [43].

ZrSiS crystalizes in a nonsymmorphic space group *P*4/*nmm* (No. 129) [Fig. 1(a)]. The bulk material can be regarded as a stack of two-dimensional (2D) monolayers. For the monolayer the nonsymmorphic symmetries lead to unavoidable band crossings from the BZ center Γ to the BZ boundary [29]. The band crossings are protected by the glide mirror symmetry $\bar{M}_z$, forming a nodal ring enclosing Γ. Owing to weak interlayer coupling, the electronic structure of bulk ZrSiS is quasi-2D. The band crossings are still present on arbitrary $k_x$-$k_y$ planes of the 3D BZ, but they are protected only on the horizontal and vertical high-symmetry planes, constituting 3D net in the BZ, as illustrated in Fig. 1(b). On the other hand, the glide and screw symmetries protect the band degeneracy at X and M points in the 2D BZ of monolayer. In the 3D case, the protected symmetries are preserved on the high-symmetry lines X-R and M-A, forming the extended nodal lines along them [31,34]. The latter is stable even in the presence of SOC, whereas the former is gapped when SOC is included, driving it to a weak topological insulator phase [29]. However, the gaps are negligible, e.g. 10-20 meV in ZrSiS [31-34], because of little hybridization between the Zr 4*d* and Si 3*p* orbitals at the band crossings [29].

As mentioned above, the electronic structure of ZrSiS has been intensively studied by VUV ARPES [31-34]. Figures 1(f), 1(j), and 1(k) show typical FSs and band dispersions near $\bar{X}$ recorded on the (001) cleavage surface of ZrSiS with VUV lights. These results are inconsistent with the calculated bulk electronic structure in Figs. 1(e) and 1(i) but well reproduced by our slab calculations with spectral weights from the topmost unit cell [Figs. 1(d) and 1(g)]. This indicates that the VUV ARPES data mainly reflect the electronic structure of the topmost unit cell because of the short mean free path of photoelectrons excited by VUV lights. The most remarkable discrepancy between the surface and bulk states is the electron FS pocket around $\bar{X}$. The dangling bonds of Zr 4*d* orbitals on the cleavage surface dramatically shift down the electron band at $\bar{X}$, leading to appearance of the electron pocket in the surface states [30].

To probe the intrinsic electronic structure of the nodal lines in the bulk states, we use soft X-ray to increase the mean free path of excited photoelectrons and therefore intrinsic definition of the out-of-plane electron momentum $k_z$ [44]. In Fig. 1(h), the FS pattern recorded with soft X-ray exhibits a diamond-shaped ring on the $k_x$-$k_y$ plane. Figure 1(c) shows that the observed FS is formed by Dirac-like band crossings at $E_F$ along a 1D closed loop, which is crucial evidence of a nodal-line FS. The soft X-ray

ARPES data in Figs. 1(l) and 1(m) exhibit a Dirac-like band crossing along Γ-X but a 'Λ'-shaped band dispersion along X-M. They are in sharp contrast with the band dispersions in the VUV ARPES data [Figs. 1(j) and 1(k)], but well consistent with the calculated bulk electronic structure at $k_z$ = 0 [Fig. 1(e)]. The excellent consistency indicates that the soft X-ray ARPES can probe the bulk states of ZrSiS without any interference from surface states.

We then use soft X-ray ARPES to systematically investigate the electronic structure related to the nodal lines. As seen in Fig. 2(a), the crossing point observed in Fig. 1(l) periodically oscillates between -0.7 eV at X and -0.5 eV at R, forming the nodal line extending along X-R across the bulk BZ. Figures 2(c) and 2(d) show the evolution of band dispersions near X in the $k_x$-$k_y$ plane. When sliding away from Γ-X, the Dirac-like band crossing is maintained but the energy of the crossing point is shifted downward [Fig. 2(c)]. The crossing points form the nearly doubly degenerate 'Λ'-shaped band along X-M. When sliding away from X-M, the 'Λ'-shaped band splits into two bands, whose tops form the Dirac bands along Γ-X [Fig. 2(d)]. These band dispersions constitute a 'saddle'-shaped band structure in the plane perpendicular to the nodal line [Fig. 2(f)]. We thus call it 'saddle' nodal line to distinguish it from the common Dirac nodal line, which has a Dirac cone-like band structure [Fig. 2(g)].

The difference of the band structures leads to distinct FS topology between them. Figure 2(b) shows the stacking ARPES intensity plot at several constant energies. At the Fermi level ($E_F$), which is above the energy of saddle point (SP), there are no FSs enclosing X. Two FSs approach X when the constant energy is lowered, and touch at X at the energy of saddle point. With further lowering the constant energy, the two FSs cross each other. By contrast, in the plane perpendicular to a Dirac nodal line, an electron FS shrinks to a point and then grows into a hole FS with lowering the energy through the Dirac point (DP) [Fig. 2(g)].

Next we focus on the electronic structure of the nodal lines on the high-symmetry planes of the bulk BZ. The calculations indicate that these nodal lines lie in the vicinity of $E_F$ and constitute 3D 'cage'-like FSs [Fig. 3(m)]. Figures 3(e) and 3(f) show the measured FSs on the Γ-M-A-Z and Γ-X-R-Z planes, respectively. We clearly resolve periodic modulations along $k_z$ of the Fermi wave vectors consistent with the calculated bulk FSs in Figs. 3(a) and 3(b), which is in contrast to the non-dispersive feature along $k_z$ in previous VUV ARPES measurements [30,32]. On the Γ-M-A-Z plane, the Dirac-like band crossing is close to $E_F$ [Figs. 3(i) and 3(k)], leading to a tiny nodal-line FS in Fig. 3(e). By contrast, on the Γ-X-R-Z plane, the band crossing at $k_z = \pi$ is obviously above $E_F$ [Fig. 3(l)], and thus the nodal-line FS on this plane exhibits a sizable area [Fig. 3(f)].

Figures 3(g) and 3(h) show the measured FSs on the $k_z = 0$ and $\pi$ planes, respectively. The FS pattern at $k_z = 0$ features an almost perfect nodal ring [Fig. 3(g)], on which the band crossing is close to $E_F$ [Figs. 3(i) and 3(j)]. By contrast, the FS pattern at $k_z = \pi$ exhibits a closed chain composed of alternate electron and hole pockets [Figs. 3(d) and 3(h)]. This is because the energy of band crossing oscillates up and down through $E_F$ rounding along the nodal ring at $k_z = \pi$ [Figs. 3(k) and 3(l)].

We summarize in Fig. 3(n) the measured FSs in the 3D bulk BZ. The two nodal rings on the $k_z = 0$ and $\pi$ planes are connected by the nodal lines on the vertical Γ-M-A-Z and Γ-X-R-Z planes, constituting the 3D 'cage'-like FSs. The bands of the nodal lines disperse almost linearly without any interference of other bands within 1.5 eV below $E_F$, which is much larger as compared with the other known TSMs. As there are no FSs other than the Dirac nodal lines, the carrier compensation enforces the nodal lines lying at $E_F$. Due to the lack of particle-hole symmetry, the crossing points of the nodal lines are not exactly at the same energy level, leading to the compensated electron-hole FS pockets. These results indicate that ZrSiS is a 'clean' system, in which the charge carriers are contributed entirely by Dirac nodal-line fermions. Transport measurements under high magnetic fields have revealed exotic properties in ZrSiS, including sharp topological phase transition, enhanced correlation effect, and magnetic breakdown [45,46], which were attributed to the remarkable nodal-line band structures. Further studies on ZrSiS are highly desired to discover unique quantum phenomena associated with the Dirac nodal-line fermions.

## Acknowledgements

This work was supported by the National Natural Science Foundation of China (11622435, 11422428, 11674369, 11474330, 11774399, 11474340, 11674371, and 11234014), the Ministry of Science and Technology of China (2016YFA0401000, 2016YFA0300600, 2013CB921700, 2015CB921300, 2016YFA0302400, 2016YFA0300300 and 2017YFA0302901), and the Chinese Academy of Sciences (XDB07000000). Y.B.H. acknowledges support by the CAS Pioneer "Hundred Talents Program" (type C).

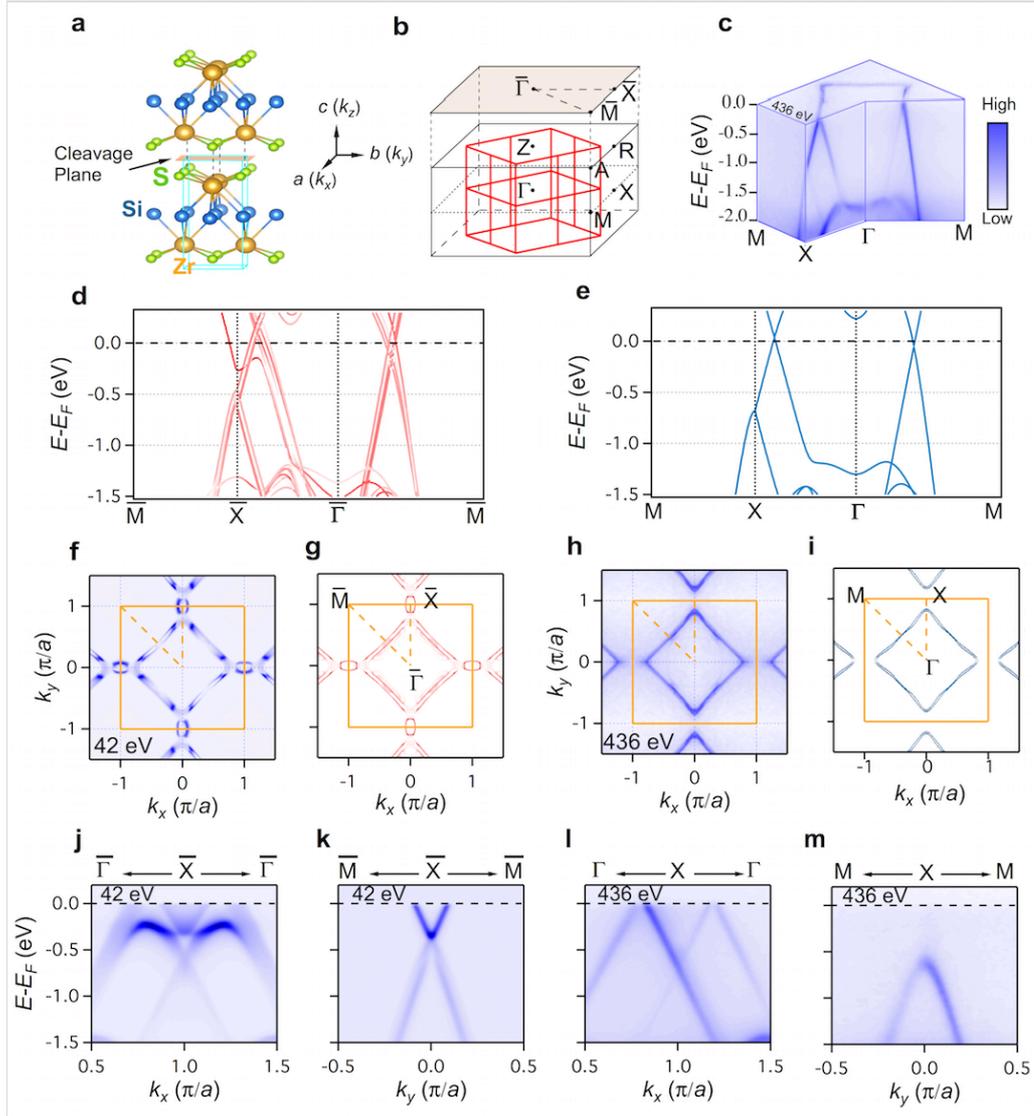

Fig. 1 (a) Crystal structure of ZrSiS. (b) 3D bulk BZ of ZrSiS and its projected (001) surface BZ. The red lines indicate the nodal lines on high-symmetry planes. (c) 3D intensity plot of ARPES spectra measured at photon energy $h\nu$ = 436 eV. (d) Calculated surface band structure. (e) Calculated bulk band structure at $k_z$ = 0. (f) FSs measured at $h\nu$ = 42 eV. (g) Calculated surface FSs. (h) FSs measured at $h\nu$ = 436 eV. (i) Calculated bulk FSs at $k_z$ = 0. (j,k) Band dispersions along $\bar{\Gamma}-\bar{X}-\bar{\Gamma}$ and $\bar{M}-\bar{X}-\bar{M}$, respectively, measured at $h\nu$ = 42 eV. (l,m) Band dispersions along Γ-X-Γ and M-X-M, respectively, measured at $h\nu$ = 436 eV.

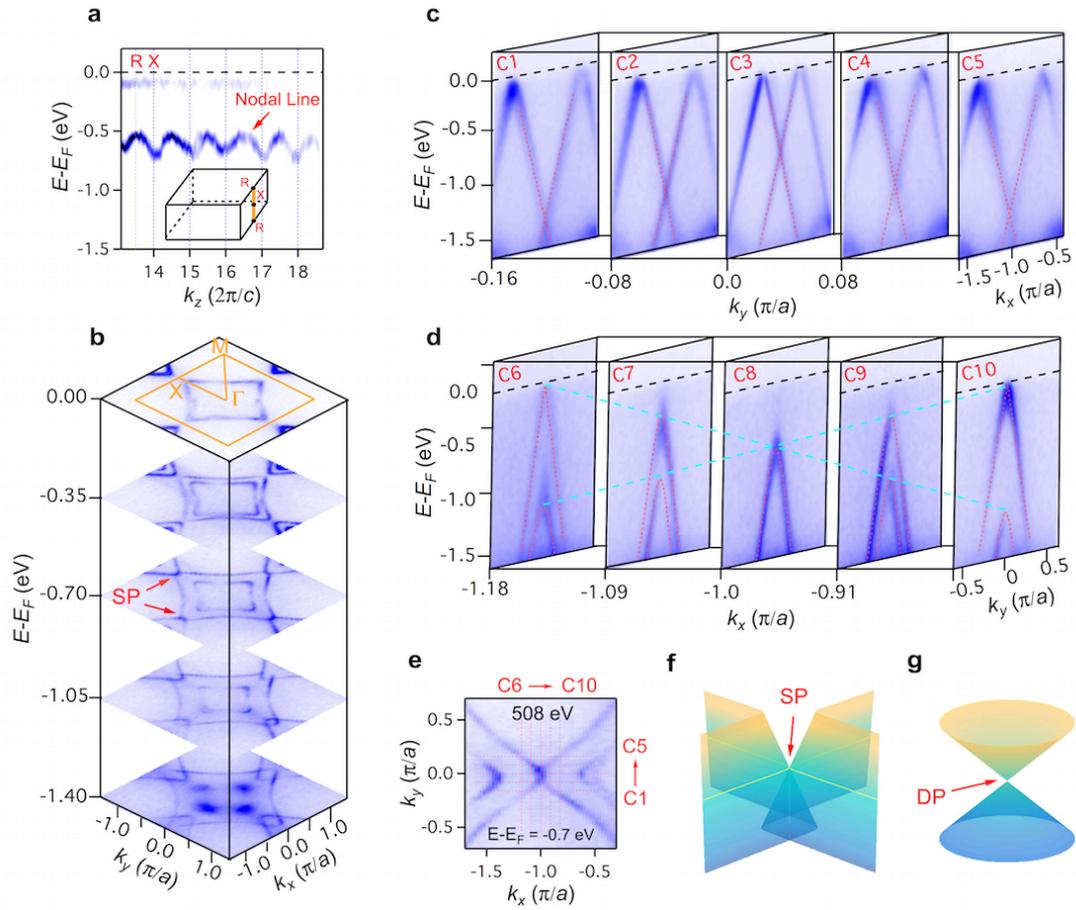

Fig. 2 (a) Intensity plot of second derivatives of the ARPES data measured in a range of $h\nu$ from 378 to 798 eV. The orange line in the inset indicates the nodal line extending along X-R. (b) Stacking ARPES intensity plot at several constant energies measured at $h\nu$ = 508 eV. (c,d) Stacking intensity plots showing band dispersions along the cuts C1-C5 and C6-C10 indicated in (e), respectively. (e) ARPES intensity plot at 0.7 eV below $E_F$. (f,g) Schematic band structures in the plane perpendicular to 'saddle' nodal line and Dirac nodal line, respectively. The yellow curves in (f) indicate the touching FSs at the energy of SP.

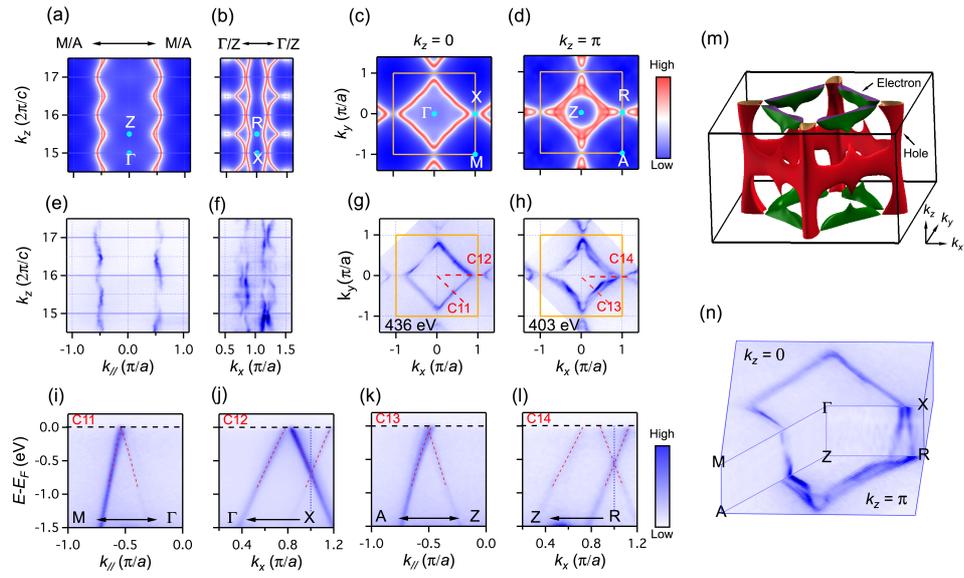

Fig. 3 (a,b) Calculated FSs in the Γ-M-A-Z and Γ-X-R-Z planes, respectively. (c,d) Calculated FSs at $k_z = 0$ and $k_z = \pi$, respectively. (e,f) Experimental FSs in the Γ-M-A-Z and Γ-X-R-Z planes, respectively. The data in (e) and (f) were measured in a range of $h\nu$ from 466 to 691 eV. (g,h) Experimental FSs at $k_z = 0$ and $k_z = \pi$, respectively. The data in (g) and (h) were measured at $h\nu$ = 436 and 403eV, respectively. (i-l) Band dispersions along the cuts C11-C14 indicated in (g) and (h), respectively. (m,n) Calculated and experimental FSs in the 3D bulk BZ.